\documentclass[reprint, amsmath,amssymb,aps]{revtex4-2}

\usepackage{graphicx}
\usepackage{dcolumn}
\usepackage{bm}
\usepackage{physics}
\usepackage{amsmath}
\usepackage{amssymb}
\usepackage{tikz}
\usetikzlibrary{quantikz2}

\begin{document}

\preprint{APS/123-QED}

\title{Suppressing Gauge Drift in Quantum Simulations with Gauge Transformations}

\author{Carter Ball}
\email{cball12@umd.edu}
\affiliation{Department of Physics, University of Maryland, College Park, MD 20742, USA}

\date{\today}

\begin{abstract}
 The simulation of quantum lattice gauge theories faces the major challenge of maintaining gauge invariance, as various errors in the simulation push the state of the system out of the physical subspace of the system's exponentially larger Hilbert space. This paper outlines a method, based off of previous work \cite{Me}, that uses gauge transformations in two ways. Firstly, the method exploits the Zeno effect by conducting frequent projections to suppress gauge drift. These projections utilize local gauge transformations to destructively interfere unphysical amplitudes via coupling to an ancillary qubit while the physical amplitudes are left untouched, up to a less than unity normalization factor. Secondly, gauge transformations are conducted throughout the time evolution of the system to hamper the speed of gauge drift. This paper demonstrates this method on a pure 1D SU$(2)$ toy model. 
\end{abstract}

\maketitle

\section{\label{sec:level1}Introduction}
Lattice gauge theories \cite{LGTintro,LGTQCD} are vital to the study of high energy physics, especially the important theories of quantum electrodynamics (QED), quantum chromodynamics (QCD), and the Standard Model. A common approach to lattice gauge theories is to use a Hamiltonian formulation, in which space is placed on a discrete lattice while time is left continuous \cite{LGTintro,HamiltonianLGT,LGToverview}. On this spatial lattice, matter fields live on the sites while gauge fields live on the links between sites. Time is kept continuous by taking the continuous time limit and employing the temporal gauge \cite{TemporalGauge}, a partial gauge that requires the time component of all gauge fields to be zero. A gauge in this context is a redundancy of degrees of freedom in the Lagrangian of a system. 

As the temporal gauge is a partial gauge, it does not fully control this redundancy; thus, there is still gauge freedom present in the spatial components of the fields. This gauge freedom is governed by a set of equations called Gauss laws, in analogy to the gauge freedoms of electromagnetic potentials in electrodynamics \cite{GaussLaw}. These Gauss laws define Gauss law operators that generate local gauge transformations at each site of the lattice. An important aspect of these lattice gauge theories is that the subspace of physically relevant states, those that are unchanged by all local gauge transformations \cite{PhysicalState}, is exponentially smaller than the total system Hilbert space. Thus, when simulating these theories, one must contend with the maintenance of gauge invariance, i.e. active consideration must be paid to keeping the system within the exponentially small subspace of gauge invariant, or physical, states. This is particularly difficult for non-abelian lattice gauge theories, such as QCD which is governed by the non-abelian gauge group SU$(3)$. 

The simulation of lattice gauge theories \cite{QuantSim,DigQuantSim1,DigQuantSim2,QuantSimUltracold,LGTsim0,LGTsim1,LGTsim2,LGTsim3,LGTsim4,LGTsim5,LGTsim6,LGTsim7,LGTsim8,LGTsim9,LGTsim10} is a major topic of research, with great potential to understand the theories underlying the structure of the physical world. These simulations contend with multiple big hurdles, including the maintenance of gauge invariance as well as the limits of classical computation. In addressing the second hurdle, much work has gone into the development of quantum computers in the hopes of this new form of computation having the power to greatly increase the productivity of quantum simulations of lattice gauge theories \cite{ComputePower1,ComputePower2,QuantCompSim1,QuantCompSim2,QuantCompSim3,QuantCompSim4,QuantCompSim5}. To that end, there are many quantum algorithms being developed to conduct these quantum simulations. 

Particular to the topic of this paper, many methods have been researched to tackle the issue of maintaining gauge invariance during quantum simulations. While in theory, if a system is initiated in a gauge invariant state and then only undergoes gauge invariant operations, then it will remain gauge invariant for the duration of the simulation. Perhaps the most important operation for simulation is time evolution; crucially, methods for gauge invariant time evolution have been developed \cite{DigQuantSim1,YukariDigQuantSim}. Unfortunately, this is often an unattainable ideal, as errors in the gauge invariance of the system's state can develop in a variety of ways, including due to necessary approximations of operations as well as quantum noise and gate errors. 

One form of quantum simulation is called analog simulation \cite{DigAnal1}, which sidesteps the difficulties of building and manipulating the system of interest by instead building a much more workable physical system that behaves like the system of interest at least in some reachable regime. Methods for analog simulations to maintain gauge invariance include satisfying gauge invariance automatically by tying it to an internal symmetry such as angular momentum \cite{InternalSym1,InternalSym2} or adding an energy penalty, effecting all unphysical states, to the system's Hamiltonian. The idea behind the energy penalty method is to craft a range of energies of the system that contains all of the physical states and no unphysical ones. There are multiple variations to this method, including adding to the Hamiltonian a term proportional to the square of the Gauss law operators \cite{EnergyPenalty1,EnergyPenalty2,EnergyPenalty3,EnergyPenalty4,EnergyPenalty5,EnergyPenalty6,EnergyPenalty7,EnergyPenalty8,EnergyPenalty9,EnergyPenalty10}, adding to the Hamilonian a term directy proportional to the Gauss law operators \cite{SingleBodyEnergyPenalty}, and a technique similar to those above but replacing the Gauss law operators with much cheaper local pseudogenerators \cite{LocalPseudoGens1,LocalPseudoGens2,LocalPseudoGens3}. Relevant to this paper's method, this local pseudogenerator method points out that term added to the Hamiltonian, when constructed of local pseudogenerators, can act as a strong projector leveraging the quantum Zeno effect, at least within some timescale, to limit the system to a quantum Zeno subspace, ie the physical subspace \cite{LocalPseudoGens3,QZsubspaces}.

The method laid out in this paper concerns a different form of quantum simulation, called digital simulation \cite{DigAnal1}, where qubits and quantum gates are used to simulate states of the system as well as operations on these states. While some methods take care to only simulate physical states, employing techniques such as solving the Gauss laws \cite{Encode1,Encode3,Encode4,Encode5,Encode6,Encode7,Encode8,LGTsim5,Twirl} or using a dual formulation \cite{Dual1,Dual2,Dual3}, this paper focuses on methods that simulate the full Hilbert space and then propose an algorithm to run within the simulation that restrain the system's state from drifting outside of the physical subspace. Such methods include living with the errors under the premise that they are reduced for smaller timesteps \cite{LiveWithIt}, a quantum control theory method of dynamical decoupling \cite{DynamicalDecoupling}, using an oracle to check for violations of gauge invariance \cite{AbelianOracle}, implementing a dynamical post-selection protocol using ancilla qubits coupled to local gauge operators \cite{Monitored1,Monitored2}, and using classical noise \cite{ZenoRandomNoise}. This last method of classical noise relies on the continuous Zeno effect to constrain the gauge drift. This is in contrast to the method to be outlined in this paper, that relies on the standard quantum Zeno effect \cite{Zeno} of frequent measurement as a method of curtailing drift. 

This paper is organized as follows. Section II outlines the method of utilizing gauge transformations in two ways to suppress gauge drift. Section III exhibits the method on a pure four-site 1D SU$(2)$ toy model with periodic boundary conditions. Section IV discusses the implementation of the method, specifically the sampling and implementation of the gauge transformations as well as any costs associated with the method. Section V concludes with a short summary and points to potential future work.

\section{Method}
For non-abelian gauge theories governed by a compact Lie gauge group, a gauge invariant, or physical, state is defined by the following set of generator equations: 
\begin{align}
    G^a_x\ket{\psi} = 0 \;\; \; \forall \;a,x
\end{align}
where $\ket{\psi}$ is a physical state and $G^a_x$ is the $a$-th generator of local gauge transformations at site x. For systems governed by unitary gauge groups, such as U$(N)$ and SU$(N)$, a local gauge transformation at a site $x$ can be written as
\begin{align}
    g_x(\alpha^a_x) = e^{i\sum\limits_a\alpha^a_xG^a_x}
\end{align}
for some set of constants $\alpha^a_x$. Combining equations (1) and (2) gives a useful second definition of a physical state as a state that is unchanged by all local gauge transformations:
\begin{align}
    g_x(\alpha^a_x)\ket{\psi} = \ket{\psi} \;\; \; \forall \;x,\alpha^a_x
\end{align}


The idea behind this paper is to leverage this relationship between physical states and gauge transformations to suppress gauge drift during the course of quantum simulations. The key insight that motivates the method of this paper concerns the action of operators of the type $e^{-it\tilde G}$ where $\tilde G$ is some operator constructed out of the gauge generators $G^a_x$. While this kind of operator can be used in gauge drift suppression schemes, as discussed below, they are expensive to construct. The function of these operators is to leave physical states untouched, as per equation (1), while affecting unphysical states in a way that reduces their prevalence in the system's overall state. The key insight of this paper, as pointed out by Lamm, Lawrence, and Yamauchi in their paper Ref.~\cite{LLY}, is that a gauge transformation does this exact thing: leaves physical states unchanged while acting on unphysical states in some way. Thus the method of this paper, designed for quantum digital simulations employing Trotterized time evolution, is to combine two methods that use $e^{-it\tilde G}$ operators for gauge drift suppression but instead use gauge transformations, which are less expensive to construct on the lattice.

The first method this paper considers is the one laid out in Ref.~\cite{LLY}, which builds their method as an alternative to the established energy penalty method of Refs.~\cite{EnergyPenalty1,EnergyPenalty2,EnergyPenalty3,EnergyPenalty4,EnergyPenalty5,EnergyPenalty6,EnergyPenalty7,EnergyPenalty8,EnergyPenalty9,EnergyPenalty10}. This method adds an energy penalty term $H_G$ to the system Hamiltonian $H_0$ such that $H_G\ket{\psi} = 0$ for every physical state and $\Vert H_G\ket{\omega}\Vert \gg 0$ for every unphysical state. Then, the system time evolves under the unitary $e^{-it(H_0+H_G)}$. As the operator $H_G$ is constructed out of gauge generators for it to act as an energy penalty, the additional term to the time evolution operator $e^{-itH_G}$ is the kind of operator this paper is interested in.

The authors of Ref.~\cite{LLY} note that for non-abelian theories, the construction of $H_G$ is difficult; thus, they propose a method to get around this by considering the effect of including the $e^{-itH_G}$ term in the time evolution operator. They point out that this term does not affect physical states but adds a phase to unphysical states, and that this is analogous to performing a gauge transformation on the system. Thus, their method dispenses with the additional energy penalty term $H_G$ in the Hamiltonian and instead conducts a gauge transformation after every Trotterized time step. Their paper shows that this method is effective at slowing down the speed of gauge drift, but it does not fully stop it. 

The method of this paper incorporates this first method into a second method to be outlined below, as they are nicely complementary. Specifically, the method below relies on the Zeno effect and thus frequent projections, so it is beneficial to slow down the speed of the gauge drift using this method of \cite{LLY} to reduce the required frequency of projections.

The second method this paper considers was developed previously by the author and their collaborator \cite{Me}, which uses the operator $e^{-itG^2}$, where $G^2 = \sum\limits_{a,x} (G^a_x)^2$, within a projection for gauge drift suppression. Below, this paper outlines this method with the crucial replacement of the $e^{-itG^2}$ operator with a random gauge transformation.

\begin{figure}[h]
    \centering
    \includegraphics[width=\linewidth]{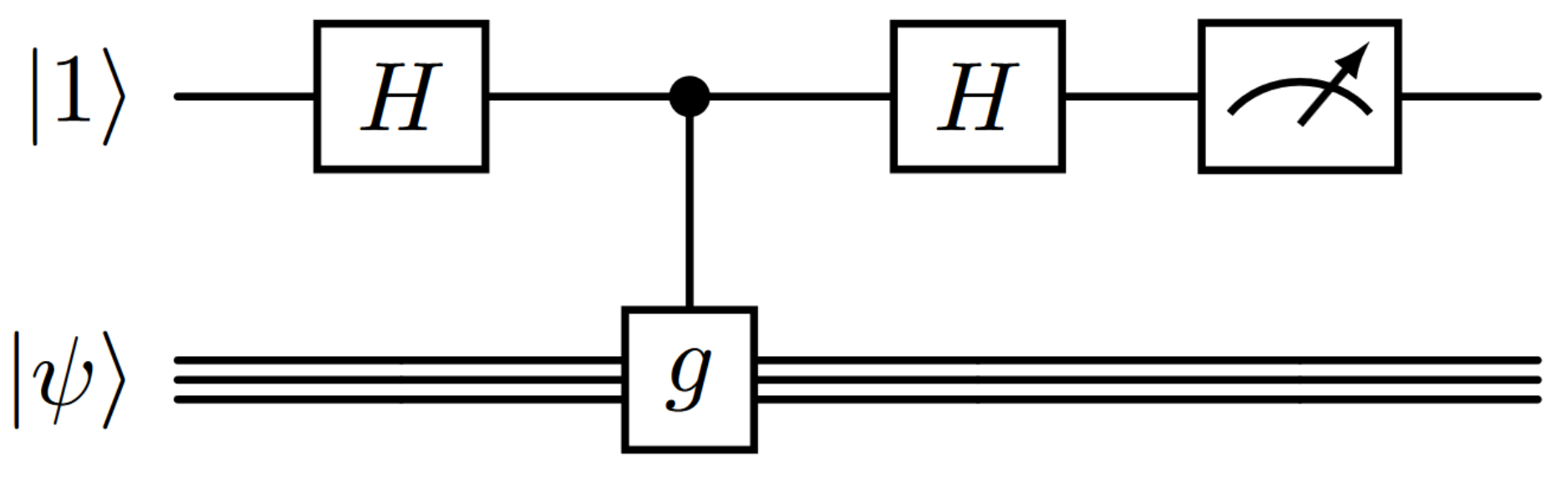}
    \caption{Circuit diagram of the projection used to suppress gauge drift}
    \label{fig:Circuit}
\end{figure}

The idea behind this method is to leverage the Zeno effect to suppress gauge drift by conducting frequent measurements which utilize gauge transformations to destructively interfere unphysical amplitudes, thus effectively projecting onto the physical subspace. The projection at the heart of this method is diagrammed in Fig.~\ref{fig:Circuit}. Note that the form of this projection is heavily inspired by the projections described by the rodeo algorithm \cite{Rodeo1,Rodeo2,Rodeo3}, an algorithm designed to construct energy eigenvectors of a given Hamiltonian.

In the projection of Fig.~\ref{fig:Circuit}, $g$ is defined as
\begin{align}
    g &= \prod\limits_x g_x 
\end{align}
Thus, $g$ is a product of a local gauge transformation on every site $x$, where each $g_x$, defined by constants $\alpha^a_x$, is chosen at random; the selection of these gauge transformations will be discussed in Sec. IV. Note that since quantum gates must be unitary operations \cite{UnitaryGates}, these gauge transformations must be unitary as well. 

This projection (minus the final measurement) is mathematically written out below:
\begin{align}
    H\Big(\ket{0}\bra{0}\otimes I + \ket{1}\bra{1}\otimes g\Big)H \ket{1}\otimes\ket\psi = \\
    H\Big(\ket{0}\bra{0}\otimes I + \ket{1}\bra{1}\otimes g\Big)\frac{\ket{0}-\ket{1}}{\sqrt{2}}\otimes\ket{\psi} = \\
    H\frac{1}{\sqrt{2}}\Big(\ket{0}\otimes\ket{\psi} - \ket{1}\otimes g\ket{\psi}\Big) = \\
    \frac{\ket{0}+\ket{1}}{2}\otimes\ket{\psi} - \frac{\ket{0}-\ket{1}}{2}\otimes\ket{\psi} = \\
    \ket{0}\otimes \frac{I-g}{2}\ket{\psi} + \ket{1}\otimes\frac{I+g}{2}\ket{\psi}
\end{align}

where the action of the Hadamard gate $H$ is understood  to act only on the ancillary qubit. 

This projection involves an ancillary qubit initialized to the $\ket{1}$ state that is first hit with a Hadamard gate to give equal amplitude, up to a sign, to its $\ket{1}$ and $\ket{0}$ components. Then a controlled-U gate entangles the ancillary qubit to the system, acting the $g$ gauge transformation on the system only for the $\ket{1}$ component of the ancillary qubit. Then, another Hadamard gate interferes the two components of the ancillary qubit, resulting in equations (8-9). Finally, the ancillary qubit is measured. A successful projection measures the qubit in the $\ket{1}$ state, otherwise the projection is said to have failed, as will become clear below.

To elucidate the effect of this projection, decompose a state into its physical, $\ket{\psi_p}$, and unphysical, $\epsilon\ket{\omega}$, parts: $\ket{\psi} = \sqrt{1-\epsilon}\ket{\psi_p} + \sqrt{\epsilon}\ket{\omega}$. Then the two components of Eq.~(9) can be better analyzed using Eq.~(3):
\begin{align}
    \frac{I-g}{2}\ket{\psi} &= \frac{\sqrt{1-\epsilon}}{2}(I-g)\ket{\psi_p} + \frac{\sqrt{\epsilon}}{2}(I-g)\ket{\omega} \\
    &= \frac{\sqrt{\epsilon}}{2}(I-g)\ket{\omega} \\
    \frac{I+g}{2}\ket{\psi} &= \frac{\sqrt{1-\epsilon}}{2}(I+g)\ket{\psi_p} + \frac{\sqrt{\epsilon}}{2}(I+g)\ket{\omega} \\
    &= \sqrt{1-\epsilon}\ket{\psi_p} + \frac{\sqrt{\epsilon}}{2}(I+g)\ket{\omega}
\end{align}
Thus, the $\ket{0}$ component of the ancillary qubit contains only unphysical parts of $\ket{\psi}$ while the $\ket{1}$ component contains all of the physical parts of $\ket{\psi}$ and only a fraction of the unphysical parts. This is why measuring the ancillary qubit in the $\ket{1}$ state is considered success and measuring in the $\ket{0}$ state is considered failure. 

To look deeper into the suppression of unphysical states in equation (13), write the unphysical component $\ket{\omega}$ in the basis of the unitary operator $g$:
\begin{align}
    \ket{\omega} &= \sum\limits_n c_n\ket{\omega_n} \\
    g\ket{\omega} &= \sum\limits_nc_ne^{i\theta_n}\ket{\omega_n} 
\end{align}
where $\ket{\omega_n}$ are eigenstates of $g$ with eigenvalues $e^{i\theta_n}$. Thus, a successful projection conducts the following (unnormalized) transformation:
\begin{align}
    \ket{\psi} &= \sqrt{1-\epsilon}\ket{\psi_p} + \sqrt{\epsilon}\sum\limits_n c_n\ket{\omega_n} \\
    &\rightarrow \sqrt{1-\epsilon}\ket{\psi_p} + \sqrt{\epsilon}\sum\limits_n\frac{1}{2}c_n (1+e^{i\theta_n})\ket{\omega_n}
\end{align}
Note that while the magnitude of the system's unphysical amplitude, characterized by $\epsilon$, affects a projection's probability of success, it does not affect the suppression mechanism of each unphysical state $\ket{\omega_n}$ picking up a multiplicative factor $\frac{1}{2}(1+e^{i\theta_n})$ (which has a magnitude $\leq 1$). The average magnitude squared of this suppression factor is
\begin{align}
    \Big\langle\Big|\frac{1}{2}(1+e^{i\theta})\Big|^2\Big\rangle &= \frac{1}{2\pi}\int\limits_0^{2\pi}d\theta \; \Big|\frac{1}{2}(1+e^{i\theta})\Big|^2 \\
    &= \frac{1}{2}
\end{align}
While \textit{a priori} there is nothing to say that the $\theta_n$ will always be uniformly distributed, as is the implicit assumption of Eq.~(18), it is a good rule of thumb that a successful projection reduces the unphysical probability density by roughly a factor of $\frac{1}{2}$, as is borne out by tests run on a SU$(2)$ toy model (see Sec.~\ref{Perf}). 

With the projection of this method outlined, it remains to be discussed the role of the Zeno effect. To that end, let a system with Hamiltonian $H$ start in the initial state $\ket{\psi_0}$ and consider the initial state amplitude $A(t)$ as well as the probability of finding the system in its initial state after a time $t$ \cite{Zeno}: 
\begin{align}
    p(t) = |A(t)|^2 = \big|\bra{\psi_0}e^{-iHt}\ket{\psi_0}\big|^2
\end{align}
Now, for short times $e^{-iHt} \simeq 1 - iHt - \frac{1}{2}H^2t^2$. Applying this expansion to the amplitude and probability gives, for short times,
\begin{align}
    A(t) &\simeq 1 - it\bra{\psi_0}H\ket{\psi_0} - \frac{1}{2}t^2 \bra{\psi_0}H^2\ket{\psi_0} \\
    p(t) &\simeq 1 - t^2\left(\bra{\psi_0}H^2\ket{\psi_0} - \bra{\psi_0}H\ket{\psi_0}^2\right)
\end{align}
This is the crux of the quantum Zeno effect: for short times the initial state amplitude decreases linearly, to leading order, while the probability decreases quadratically due to the cancellation of linear terms. Thus, if the system is time evolved for only a short time and then measured, there is a high probability of finding the system in its initial state. Note here that while the above case deals with an individual state, the quantum Zeno effect works with subspaces as well. Thus, if frequent enough measurements are taken throughout the course of a simulation, the system will to good approximation remain within the desired state or subspace for the duration of the simulation. For the purposes of this paper the desired subspace is of course the physical subspace of the system's Hilbert space.

Previous work by the author and their collaborator \cite{Me} explicitly sketches this basic scenario out: consider a system that will evolve for a total time $T$ during which $N$ projections take place, spaced evenly. While the projections outlined above are not perfect, for the sake of this exercise they will be treated as such. Furthermore, say that $r$ is the rate at which the system's state accumulates amplitude outside of the desired space, i.e. unphysical amplitudes. Then the probability of a failed projection is $(\frac{rT}{N})^2$. Then the probabilty of the full simulation completing with 0 failures is
\begin{align}
    P_s = \left(1-\frac{r^2T^2}{N^2}\right)^N \approx 1 - \frac{r^2T^2}{N}
\end{align}
in the limit of large $N$ and $1 - P_s \gg 1$. Notably, this is close to one when $N \gtrsim r^2T^2$. Thus, frequent enough projections can keep the system within the physical subspace for the entirety of the simulation. 

At this point, failed projections must be addressed. If a projection does fail, then the whole simulation must be restarted. Obviously, this comes at great cost and so should be avoided if at all possible. A first solution would be to do a projection after every single Trotterized time step. As this gives the system very little time to drift out of the physical subspace, this greatly suppresses the chance of a failed projection; however, this is costly. Therefore, an optimization problem balancing the cost of projections and the risk of a failed projection must be addressed.

Recall that the method of this paper has two components: one being the frequent projections to leverage the Zeno effect and the other being the use of gauge transformations after every Trotterized time step. As discussed above, this second component slows down the rate of the gauge drift, which in turn helps address this optimization problem since slower gauge drift reduces the necessary frequency projections. 

Furthermore, a simple calibration phase can be used to tune the frequency of projections to an optimal value. This calibration phase is best used for simulations that plan to be run many times. This phase begins by selecting an initial frequency of projections based off of knowledge about the system and the cost of projections. With the initial frequency chosen, run the simulation a handful of times till failure. With multiple runs of this test, the simulators should get a good idea of if the initial frequency of projections will work for the parameters of their simulation; if not, adjust the frequency accordingly and go again. In this way, a workable frequency of projections can be found fairly simply.  

In summary, the method of this paper is to leverage the Zeno effect to suppress gauge drift by performing frequent projections (diagrammed in Fig.~\ref{fig:Circuit}) while also performing a random gauge transformation after every Trotterized time step of the simulation.

\section{Performance}\label{Perf}
\begin{figure}[h]
    \centering
    \includegraphics[width=0.5\linewidth]{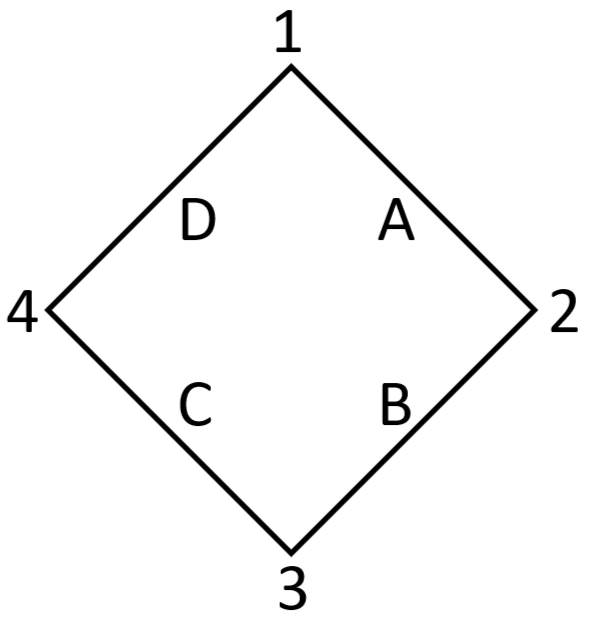}
    \caption{The diagram of the 1D toy model with periodic boundary conditions, four sites (labelled 1-4), and four links (labelled A-D).}
    \label{fig:ToyModel}
\end{figure}
In order to see the method of this paper in action, consider a 1D pure SU$(2)$ toy model with 4 sites, taking a Kogut-Susskind lattice approach. Periodic boundary conditions were used to avoid large edge effects one might see for such a small system and instead focus on bulk behavior. Fig.~\ref{fig:ToyModel} diagrams this toy model, representing it as a diamond shape to clarify that this is a one-dimensional system with periodic boundary conditions instead of a singular plaquette. The Hamiltonian for this toy model is 
\begin{align}
    H = \lambda\sum\limits_{i=A}^D  E^2_i
\end{align}
where $E^2_i$ is the quadratic Casimir operator of the $i$th link and $\lambda$ is a constant, set to 1 for convenience. Note that the toy model is simulated in a representation basis with a cutoff $j_{max}=1/2$ imposed. This $j$-cutoff is an analogy to angular momentum and limits the allowed eigenvalues of the quadratic Casimir operator $E^2_i$ which have the form $j(j+1)$. Thus, this cutoff  limits the available energy levels of each link. This does not affect the work of this paper as limiting a system to its lower energy levels still allows for the study of gauge drift and its suppression.

Each site has 3 generators of local gauge transformations at that site; as an example
\begin{align}
    G^x_1 &= L^x_A + R^x_D \\
    G^y_1 &= L^y_A + R^y_D \\
    G^z_1 &= L^z_A + R^z_D 
\end{align}
where the superscripts $x,y,z$ reference the three Pauli matrices and the $L_i$ ($R_i$) operator is the left (right) electric field operator on the $i$th link. This means a local gauge transformation at site $1$ can be written
\begin{align}
    g_1 &= e^{i(\alpha^x_1G^x_1 + \alpha^y_1G^y_1 + \alpha^z_1G^z_1)} \\
    &= e^{i\vec\alpha_1 \cdot\vec G_1}
\end{align}
where the $\alpha^i_1$'s are constants and we have defined $\vec G_i = (G^x_i,G^y_i,G^z_i)$ for convenience. With this in hand, any gauge transformation on the whole system can then be written
\begin{align}
    g = g_1g_2g_3g_4 = e^{i\vec\alpha_1 \cdot\vec G_1}e^{i\vec\alpha_2 \cdot\vec G_2}e^{i\vec\alpha_3 \cdot\vec G_3}e^{i\vec\alpha_4 \cdot\vec G_4}
\end{align}

As the goal of this paper is to study the suppression of gauge drift, an artificial drift operator $D$ is used to simulate gauge drift. For concreteness, it is helpful to label the toy model's two physical states as $\ket{0}$ and $\ket{1}$ and all other, unphysical states $\ket{\omega_i}$. The idea behind the drift operator is to unitarily connect each physical state to a randomly selected unphysical state; call them $\ket{\omega_a}$ and $\ket{\omega_b}$. Thus, the drift operator was constructed as following:


\begin{align}
    &D(\epsilon,\omega_a,\omega_b) = \sum\limits_{i\neq a,b}\ket{\omega_i}\bra{\omega_i} + \\
    &+\sqrt{1-\epsilon^2}\Big(\ket{0}\bra{0}+\ket{1}\bra{1}+\ket{\omega_a}\bra{\omega_a}+\ket{\omega_b}\bra{\omega_b}\Big) + \nonumber\\ &+\epsilon\Big(\ket{0}\bra{\omega_a}-\ket{\omega_a}\bra{0}+\ket{1}\bra{\omega_b}-\ket{\omega_b}\bra{1}\Big)\nonumber
\end{align}
where $\epsilon$ parametrizes the strength of the drift operator. 

As can be seen, the drift operator exchanges amplitude between $\ket{0}$ and $\ket{\omega_a}$ as well as between $\ket{1}$ and $\ket{\omega_b}$; for the mostly physical states that this drift operator will act on, this means the drift operator reduces the physical probability density and increases the unphysical probability density. When it comes to implementation, the drift operator is used on the toy model's state right before every time evolution. Crucially, for every instance of the drift operator, new unphysical states are randomly selected so as to not preference any in particular. 

The final aspect of studying the suppression of gauge drift is to quantify the ``physicalness", or equivalently ``unphysicalness", of a given state. This can be done with the following operator:
\begin{align}
    G^2 = \sum\limits_{a,x} (G^a_x)^2
\end{align}
This operator, discussed above in Sec.~II as well as in Ref.~\cite{Me}, is defined such that the ground state eigenspace, with eigenvalue 0, is exactly all physical states and no unphysical states. Thus a given state $\ket{\psi(t)}$ can be evaluated on its ``physicalness" by considering the expectation value $\bra{\psi(t)}G^2\ket{\psi(t)}$. If this expectation value is 0, the state is fully physical; the further from 0 it gets the more unphysical the state is said to be. Note that by construction $G^2$ is positive definite, so $\expval{G^2} \geq 0$ for any state.

The simulations run for this paper employed the Trotterization technique, wherein the system was time evolved by repeated use of the unitary $e^{-iH\Delta t}D(\epsilon)$ for $\Delta t = 0.01$ time units and $\epsilon = 0.01$. After each instance of this unitary there was an opportunity for a gauge drift mitigation technique to be used, such as a random gauge transformation for a projection; the choices concerning which method is used and how frequently they are used define a simulation's overall mitigation scheme.  

Every run of the algorithm laid out in this paper starts by initializing the system in a fully physical state. As there are two physical states for this system (one where all links are $j=0$ and another where all links are $j=1/2$ with the appropriate $m_j$ values at the ends), this initialization amounts to randomly generating an amplitude for one physical state, which then determines the amplitude (up to a phase, chosen to make the amplitude real and positive) of the other physical state, given the constraint that the initial state is normalized. 

\begin{figure}[h]
    \centering
    \includegraphics[width=\linewidth]{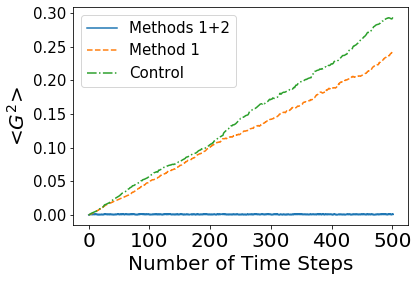}
    \caption{Results from three simulation runs: a control run, dashed-dot line, without any gauge drift suppression methods; a run, dashed line, using method 1 (conducting a gauge transformation after every time step); and a run, solid line, using both method 1 and  method 2 (conducting a projection after every time step).}
    \label{fig:DriftA}
\end{figure}
\begin{figure}[h]
    \centering
    \includegraphics[width=\linewidth]{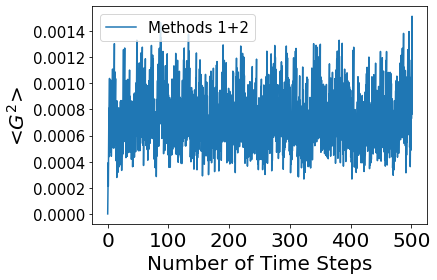}
    \caption{A closer look at the run from Fig.~\ref{fig:DriftA} utilizing both methods of gauge drift suppression.}
    \label{fig:DriftB}
\end{figure}

As a first test, three simulations were run on a classical computer for 500 time steps each, calculating $\expval{G^2}$ after every time step. These three simulations were chosen to demonstrate the basic performances of the methods used; see Figure \ref{fig:DriftA}. The first run, a control run, did not employ any suppression techniques after each Trotterized time step (dashed-dot line); the second run implemented the first method of gauge drift suppression discussed in this paper, i.e. the use of a random gauge transformation after each time step (dashed line); the third run used both this first method with gauge transformations as well as the second method of performing the projection diagrammed in Fig.~\ref{fig:Circuit} after every time step (solid line). Figure \ref{fig:DriftB} plots only this third run to highlight its performance. 

Figures \ref{fig:DriftA} and \ref{fig:DriftB} show that the method of this paper is quite effective at suppressing gauge drift for this toy model. As expected, leaving the system without any gauge drift suppression technique (control) allows for unphysical amplitudes to build up over time, represented by a growing $\expval{G^2}$ value. Furthermore, the use of method 1 slows down this growth but does not bound it. Finally, employing both methods bounds the growth of unphysical amplitudes to a small region close to the physical subspace, represented by $\expval{G^2} = 0$. Thus, Fig.~\ref{fig:DriftB} shows that the system remains highly physical throughout its simulation when both methods of this paper are implemented.


Looking further into the method of this paper, it is useful, particularly during the calibration phase, to consider the average suppression factor of a projection. To that end, a test was run wherein the system was initialized to a random physical state 50 times and then allowed to run for 101 time steps. Every run was simulated with a projection conducted after each Trotterized time step; additionally, the expectation value $\expval{G^2}$ was calculated before and after each projection. A suppression factor for each projection was then calculated as $\expval{G^2}_{after}/\expval{G^2}_{before}$. The average of all 101 suppression factors from all 50 initializations (for a total of 5,050 suppression factors) was calculated to be $0.512 \pm 0.063$. Thus, the working estimate of a $50\%$ suppression factor posited in Sec.~II is borne out in the data. 

Now, this method relies fundamentally on performing random gauge transformations frequently throughout the evolution of a system, which can be a very large number of gauge transformations that must be sampled and performed. Thus, it is worthwhile to explore the possibility of instead using a small set of gauge transformations repeatedly throughout a run of a simulation.

To that end, simulations were run with the reuse of gauge transformations to study how this might affect the effectiveness of the gauge suppression method. For the purposes of direct comparison, each of the subsequent simulations followed the same general simulation pattern, consisting of four steps: 
\begin{enumerate}
    \item Time step
    \item Gauge transformation (method 1)
    \item Time step
    \item Projection (method 2)
\end{enumerate}
Crucially, recall that a gauge transformation is used in step 4 as well as in step 2; see the projection's diagram in Fig.~\ref{fig:Circuit}. For each simulation, this pattern was repeated 250 times, for a total of 500 time steps. Furthermore, $\expval{G^2}$ was calculated after each of the four steps listed above. 

With this basic pattern, four different methods of simulation were created. Each method of simulation was run 200 times, and then the $\expval{G^2}$ values from all 200 runs were averaged; these averages are shown in Table \ref{Table}. The first method, a control, randomly generated every gauge transformation used throughout the simulation, in both steps 2 and 4. For the remaining three methods of simulation, a simulation's initialization would include generating two gauge transformations, call them $g_1$ and $g_2$, to be used repeatedly throughout said simulation. The second method of this test, called ``Alternate time step $g$'s" in Table \ref{Table}, alternated between using $g_1$ and using $g_2$ for step 2, while the gauge transformations in step 4 were randomly generated. The third method, called ``Alternate projection $g$'s" in Table \ref{Table}, did the opposite: the gauge transformations in step 2 were randomly generated while step 4 alternatively used $g_1$ and $g_2$. The fourth method, called ``Alternate all $g$'s" in Table \ref{Table}, used $g_1$ and $g_2$ as the only gauge transformations throughout the simulation; more specifically, the first step 2 used $g_1$, then the first step 4 used $g_2$, followed by the second step 2 using $g_2$ and the second step 4 using $g_1$. This pattern repeated throughout each simulation.


\begin{table}[h]
\centering
\begin{tabular}{||c|c||}
    \hline
    Method & Average $\expval{G^2}$ value \\
    \hline\hline
    Control (fully random $g$'s) &  $(14.8 \pm 4.2) \times 10^{-4}$ \\
    \hline
    Alternate time step $g$'s & $(14.9 \pm 4.2) \times 10^{-4}$ \\
    \hline
    Alternate projection $g$'s & $(14.6 \pm 4.1) \times 10^{-4}$ \\
    \hline
    Alternate all $g$'s & $(17.2 \pm 4.3) \times 10^{-4}$ \\
    \hline
\end{tabular}
\caption{The average $\expval{G^2}$ value for four different methods of reusing gauge transformations throughout a simulation run}
\label{Table}
\end{table}

Now, the method of this paper posits that the use of a singular gauge transformation is not going to provide good suppression for every unphysical state, but that one instance of a gauge transformation can provide good suppression cumulatively. Furthermore, this method poses that multiple different gauge transformations used throughout a simulation can provide strong suppression, the assumption being that if one gauge transformation fails to adequately suppress a certain unphysical state then the next gauge transformation(s) can pick up the slack. Thus, the method of randomly generating gauge transformations was used as the baseline method to avoid any suppression-affecting biases. 

The results of Table \ref{Table}, however, show that this might not be too significant a concern. These results demonstrate, for this toy model, that reusing gauge transformations does not have an outsized effect on the performance of the method. Thus, while the toy model in this paper is quite small and trivial, this data provides evidence suggesting that it is unnecessary to fully randomly generate a new gauge transformation every time one is needed and that perhaps a less arduous implementation scheme, such as reusing a small number of gauge transformations, can be effective. 

It should be noted that this concept of reusing gauge transformations does have a lower limit. Even for this toy model, using only one gauge transformation for the entirety of a simulation destroyed the efficacy of the method. At least for this toy model, this was because any given gauge transformation has at least some unphysical states with an eigenvalue close to 1, so they don't receive adequate gauge drift suppression to counter the drift coming from the drift operator $D$ used in conjunction with each time step. While this turned out to be the case for only using one gauge transformation, this did not damage the efficacy the case of using two gauge transformations throughout as one gauge transformation could cover for the weaknesses of the other.

\section{Implementation}\label{Imp}
The method of this paper relies heavily on sampling and performing gauge transformations on the system of interest. Thus, it is prudent to explicitly discuss the implementation of these actions. When it comes to sampling the gauge transformation, especially if gauge transformations are being reused as discussed above, it might be best to sample them on a classical computer and then hard-code them into the quantum circuit before beginning any simulation \cite{LLY}. This runs into the risk of `weighing down' the circuit with too much information, depending on the required number of gauge transformations to hard-code. 

Another option, discussed in both \cite{LLY} and \cite{Me}, is to use an ancillary `clock' register that simply counts the number of gauge transformations that have been performed during the simulation. This clock register feeds its value into a random circuit of the type in Ref.~\cite{Short} that will output a gauge transformation, storing it in an ancillary G-register. This method clearly comes with qubit costs, namely the additional clock and G-registers. Note however that these costs do not scale with the size of the system. 

Another cost of this method is the depth of the random circuit that generates gauge transformations. One thing to note here is that Sec.~III shows the repeated use of only two gauge transformations was effective; this implies that the efficacy of this paper's method is not highly sensitive to the level of fairness of the gauge transformation sampling. Thus, a shorter random circuit, which intuitively samples less fairly than longer random circuits, can be used without really damaging the efficacy of the method. Furthermore, Ref.~\cite{LLY} argues that unfair sampling can be made more fair by conducting multiple unfairly sampled gauge transformations in a row. Thus, it is quite probable that a relatively short random circuit would suffice. 

When it comes to the actual implementation of the gauge transformations on the system in the course of the simulation, this “requires about as many G-multiplication gates as the original quantum simulation did” \cite{LLY}.

One final point to discuss is the qubit cost of the projections discussed above. Crucially, note that each measurement only requires one ancillary qubit. Thus, when mid-circuit measurements are a feature of the simulation, only one ancillary qubit is required, to be used for each projection. If this is not the case, then the qubit cost for the full simulation is equal to the total number of measurements to be performed.

\section{Conclusion}
This work outlines a method of suppressing gauge drift with gauge transformations during quantum simulation of unitary lattice gauge theories. This method uses frequent projections to leverage the Zeno effect to achieve this suppression; furthermore, it includes the use of gauge transformations in the course the time evolution of the system that hamper the gauge drift. This paper shows this method to be effective for a toy model of a four-site pure 1D SU$(2)$ system with periodic boundary conditions. Future work aims to test the method out on larger systems as well as systems with matter fields. Furthermore, deeper research into the reuse of gauge transformations and the required level of fairness of their sampling is warranted.

\begin{acknowledgments}
Useful conversations with Thomas D. Cohen are gratefully acknowledged, as well as the work done with him previously that vitally informed this work. This work was supported in part by the U.S. Department of Energy, Office of Nuclear Physics under Award Number(s) DE-SC0021143 and DE-FG02- 93ER40762.
\end{acknowledgments}

\bibliography{biblio.bib}
\end{document}